\documentclass[lettersize, journal]{IEEEtran}

\IEEEoverridecommandlockouts
\usepackage{bm}
\usepackage{cite}
\usepackage{url}
\usepackage{float}
\usepackage{amsmath}
\usepackage{color}
\usepackage{amssymb,amsthm,graphicx,subfigure,cite,color,enumerate,algorithm,algorithmic}
\usepackage{diagbox}
\usepackage[table]{xcolor}
\usepackage{booktabs}
\usepackage{stfloats}
\usepackage{CJK}
\usepackage{subfigure}
\usepackage{balance}
\usepackage{footnote}
\usepackage{multirow}
\makesavenoteenv{align}
\newtheorem{lemma}{Lemma}

\makeatletter
\newenvironment{breakablealgorithm}
  {% \begin{breakablealgorithm}
   \begin{center}
     \refstepcounter{algorithm}% New algorithm
     \hrule height.8pt depth0pt \kern2pt% \
     \renewcommand{\caption}[2][\relax]{% Make a new \caption
       {\raggedright\textbf{\ALG@name~\thealgorithm} ##2\par}%
       \ifx\relax##1\relax % #1 is \relax
         \addcontentsline{loa}{algorithm}{\protect\numberline{\thealgorithm}##2}%
       \else % #1 is not \relax
         \addcontentsline{loa}{algorithm}{\protect\numberline{\thealgorithm}##1}%
       \fi
       \kern2pt\hrule\kern2pt
     }
  }{% \end{breakablealgorithm}
     \kern2pt\hrule\relax%
   \end{center}
  }
\makeatother
\allowdisplaybreaks[4]
\def\BibTeX{{\rm B\kern-.05em{\sc i\kern-.025em b}\kern-.08em
    T\kern-.1667em\lower.7ex\hbox{E}\kern-.125emX}}
\usepackage{balance}

\begin{document}
\title{Hierarchical Codebook Design for Near-Field MmWave MIMO Communications Systems }
\author{Jiawei~Chen, Feifei Gao, Mengnan Jian and Wanmai Yuan
\thanks{J. Chen and F. Gao are with Beijing National Research Center for Information Science and Technology (BNRist),
Department of Automation, Institute for Artificial Intelligence, Tsinghua University (THUAI), Beijing 100084, China (e-mail:
chenjiaw20@mails.tsinghua.edu.cn; feifeigao@ieee.org).}
\thanks{M. Jian is with the State Key Laboratory of Mobile Network and Mobile Multimedia Technology, Shenzhen 518055, China, and also with the Algorithm Department, Wireless Product R\&D Institute, ZTE Corporation, Shenzhen 518057, China (e-mail: jian.mengnan@zte.com.cn).}
\thanks{W. Yuan is with Information Science Academy of CETC. (email: yuanwanmai7@163.com)}
}

\maketitle

\begin{abstract}
  Communications system with analog or hybrid analog/digital architectures usually relies on a pre-defined codebook to perform beamforming.
  With the increase in the size of the antenna array, the characteristics of the spherical wavefront in the near-field situation are not negligible.
  Therefore, it is necessary to design a codebook that is adaptive to near-field scenarios.
  In this letter, we investigate the hierarchical codebook design method in the near-field situation.
  We develop a steering beam gain calculation method and design the lower-layer codebook to satisfy the coverage of the Fresnel region.
  For the upper-layer codebook, we propose beam rotation and beam relocation methods to place an arbitrary beam pattern at target locations.
  The simulation results show the superiority of the proposed near-field hierarchical codebook design.
\end{abstract}


\section{Introduction}\label{sec:intro}
In the next generation communication system, beamforming techniques with large antenna arrays are necessary to help compensate for the severe pathloss caused by ultra-high frequency in mmWave or Terahertz frequency band \cite{Terasurvey, beam5g}.
Due to the massive antenna array size as well as the high power consumption and cost of the radio-frequency (RF) chain, the fully-digital beamforming is hard to be implemented \cite{costdigital}.
Instead, analog beamforming with a single RF chain and the constant amplitude constraint is more practical, which usually relies on a pre-defined codebook.

There have been extensive researches in the codebook design with different metrics such as the coverage in a particular domain \cite{dft,commonCB} or the reduction of searching steps \cite{cbdeact,cbbwmss}.
The widely implemented Discrete Fourier Transform (DFT) codebook \cite{dft} in uniform linear array (ULA) aims to provide coverage of angular domain, while the authors in \cite{commonCB} propose a codebook design for uniform planar array (UPA) that covers block areas in azimuth and elevation domain.
To reduce the beam searching steps, other researches focus on hierarchical structure design, where only a subset of beams needs to be searched at each layer.
The authors of \cite{cbdeact} proposed a hierarchical codebook design that generates wide beams by turning off antennas.
In \cite{cbbwmss}, a hierarchical codebook that jointly utilizes sub-array and deactivation techniques (BWM-SS) is developed.

On the other hand, the large size of antenna arrays in mmWave or Terahertz frequency band would make the characteristics of the spherical wavefront non-negligible.
Therefore, the codebook designed for far-field scenario \cite{dft,commonCB,cbdeact,cbbwmss} may suffer from performance loss, and near-field characteristics should be taken into consideration.
In \cite{channelest}, a channel representation method is investigated in the polar domain, where the channels are sampled on a distance ring.
The authors of \cite{nearhier1} develop a near-field phase-shift codebook that is suitable for Reconfigurable Intelligent Surfaces (RISs).
% In \cite{nearhier2}, a hierarchical codebook design for RIS in the near-field situation based on sampled points is proposed.
However, the codebook in \cite{nearhier1} is designed for RISs and does not consider the coverage for the whole polar domain.

In this paper, we investigate the codebook design of a ULA that is suitable for the near-field situation.
The designed hierarchical codebook considers a spherical wavefront and provides coverage for the whole Fresnel region.
Specifically, we separately design the lower-layer codebook that provides coverage for the polar domain and the upper-layer codebook that aims to reduce the overhead of beam selection.
For the lower-layer codebook, we utilize the steering beam vector to provide beamforming and investigate the beam gain of the steering beam pattern in the polar domain.
As for the upper-layer codebook, we propose beam rotation and beam relocation methods to place the initial beam pattern at target locations.
The simulation results show the effectiveness of the proposed upper-layer and lower-layer codebook design methods.

%\emph{Notations:} In this paper, scalars are denoted by lower-case letters $a$, vectors are denoted by boldface letters $\mathbf{a}$, matrixes are denoted by capital letters $\mathbf{A}$, and sets are denoted by $\mathcal{A}$.
%The $l$- norm and absolute operators are denoted by $\|\cdot\|_l$ and $|\cdot|$, respectively.
%Transpose and Hermitian operators are denoted by $(\cdot)^T $and $(\cdot)^H$.
%the $\circ$ denotes Hadamard product.
%Moreover, $\mathbf{a}[i]$ denotes the $i$-th element of vector $\mathbf{a}$.
%$\mathbb{R}^M$ and $\mathbb{C}^M$ denote the M-dimensional real and complex spaces, respectively.

\section{System Model}\label{sec:sys}
We consider a mmWave communications system where a transmitter performs analog beamforming for a receiver.
The transmitter equips with a single radio-frequency (RF) chain and a uniform linear array (ULA) of $n_w$ antennas, while the receiver equips with a single antenna.
The analog beamforming is realized by adjusting the phase shifter attached to each antenna, while the amplitude of each element is constant.
The received signal at the receiver can be expressed as:
\begin{equation}
  y = \mathbf{h}^H \mathbf{w} s + n,
\end{equation}
where $\mathbf{h} \in \mathbb{C} ^{n_w\times 1}$ is the channel vector between the transmitter and the receiver, $s$ denotes the transmit signal that satisfies $E\{|s|^2\} = P_s$, $\mathbf{w}$ denotes the analog beamforming vector, and $\mathbf{n}\in \mathcal{CN} (0, \sigma^2 \mathbf{I})$ is the noise.
It is assumed that all receivers and scatters are located within the Fresnel region of the transmitter and thus we adopt the near-field channel model
\begin{equation} \label{eq:channel}
  \mathbf{h} = \sum_{l=1}^{L} \alpha_l \mathbf{a} (\theta_l, r_l),
\end{equation}
where $\alpha_l \in \mathbb{C}$ is the complex path component of the $l$-th path, and $L$ is the number of paths between the transmitter and the receiver.
In the near-field scenario, the spherical wave characteristic should be taken into consideration.
The steering vector applied to the near-field situation of the ULA can be expressed as
\begin{equation}
  \mathbf{a}(\theta, r) = \frac{1}{\sqrt{n_w}} \left[e^{-j\frac{2\pi}{\lambda}(r_{1} - r)}, ..., e^{-j\frac{2\pi}{\lambda}(r_{n_w} - r)}\right]^T
\end{equation}
where $\lambda$ denotes the carrier wavelength, $r$ denotes the distance between the central point and the receiver or scatter while $r_{i}$ denotes the distance between the $i$-th element of the ULA and the receiver or scatter.
According to \cite{channelest}, the distance $r_{i}$ can be approximated as $r_i = \sqrt{r^2 + \delta_i^2d^2 - 2r\theta \delta_i d} \approx r-\delta_id\theta+\frac{\delta_i^2d^2(1-\theta^2)}{2r}$, where $\delta_i=\frac{2i-n_w-1}{2}$ and $d$ denotes the antenna space of ULA.
Therefore, the normalized beam gain $g(\mathbf{w}, r, \theta)$ of codeword $\mathbf{w}$ at location $(r, \theta)$ can be calculated as
\begin{equation} \label{eq:gain}
  \begin{split}
    g(\mathbf{w}, r, \theta) &= \left|\mathbf{w}^H \mathbf{a}(\theta, r)\right| \\
    &= \left|\frac{1}{n_w} \sum_{-(n_w-1)/2}^{(n_w-1)/2} [\mathbf{w}]_n e^{j \frac{2\pi}{\lambda} (- r^{(n)})} \right| \\
    &\overset{\mathrm{(a)}}{\approx} \left|\frac{1}{n_w} \sum_{-(n_w-1)/2}^{(n_w-1)/2} [\mathbf{w}]_n e^{jn\pi \theta - j\frac{2\pi}{\lambda}
    n^2 d^2\left(\frac{1 - \theta^2}{2 r}\right)}\right|
  \end{split}
\end{equation}
where $\mathbf{w} = \frac{1}{\sqrt{n_w}} [e^{j\phi_1}, ..., e^{j\phi_n}]^T$ has constant amplitude constraint, $\phi_i $ is the phase shift of the $i$-th element of the ULA.
It is noted that the $\mathrm{(a)}$ is accurate within the Fresnel region \cite{channelest}, where the distance $r$ satisfies $r > r_{min} = 0.5\sqrt{\frac{D^3}{\lambda}}$, and $D=n_w d$ is the size of the antenna array.

The codebook-based beamforming is to choose the optimal beam vector $\mathbf{w}$ in codebook $\mathbf{W} = [\mathbf{w}_{11}, ...,  \mathbf{w}_{1n^\theta}, ...,  \mathbf{w}_{n^r1}, ..., \mathbf{w}_{n^rn^{\theta}}]^T$ that maximizes the beam gain at the receiver for given channel $\mathbf{h}$, i.e.,
\begin{equation}
  \begin{split}
    \max_{\mathbf{w} \in \mathbf{W} } \quad & |\mathbf{w}^H \mathbf{h}| \\
    s.t. \quad & |\mathbf{W}[i, k]| = \frac{1}{\sqrt{n_w}}, \forall i\in \{1, ..., n^rn^{\theta}\}, k \in \{1,..., n_w\},
  \end{split}
\end{equation}
where $n^{r}$ and $ n^{\theta}$ are the number of codewords along $r$-dimension and $\theta$-dimension in polar domain, respectively.
Based on beam gain (\ref{eq:gain}), the coverage region of beam vector $\mathbf{w}$ in near-field scenario can be defined as
\begin{equation}
  \mathcal{C}(\mathbf{w}) = \left\{(\theta, r)  \bigg| g(\mathbf{w}, \theta, r) > \rho \max_{(\hat{\theta}, \hat{r})} g(\mathbf{w}, \hat{\theta}, \hat{r})\right\}
\end{equation}
where $\rho$ is the scalar within $(0,1)$.

\section{Hierarchical Codebook Design}
\begin{figure}
  \centering
  \includegraphics[width=0.40\textwidth]{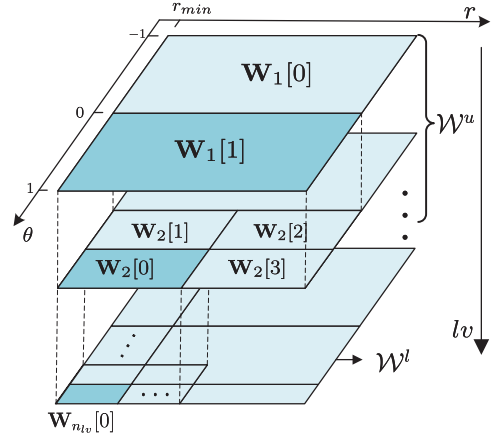}
  \caption{The illustration of hierarchical codebook in the near field situation.}
\label{fig:cblv}
\end{figure}

In this section, we provide a detailed hierarchical codebook design method that is suitable for near-field scenarios.
%The hierarchical codebook design aims to reduce the beam searching steps.
As illustrated in Fig. \ref{fig:cblv}, the hierarchical codebook $\mathcal{W} = \{\mathbf{W}_{1}, \mathbf{W}_{2}, ..., \mathbf{W}_{n_{lv}}\}$ is consisted of $n_{lv}$ levels of codebook.
The upper-layer codebooks $\mathcal{W}^{u} = \{\mathbf{W}_{1}, \mathbf{W}_{2}, ..., \mathbf{W}_{n_{lv}-1}\}$ is designed to reduce the beam searching steps, while the lower-layer codebook $\mathcal{W}^{l} = \{\mathbf{W}_{n_{lv}}\}$ aims to provide high beamforming gain for receivers in Fresnel region.

\subsection{Lower-layer Codebook Design}
%$\mathbf{W}_{n_{lv}}$
The aim of lower-layer codebook $\mathcal{W}^{l}$ is to match the strongest multipath components (MPCs) in the channel and provide high beamforming gain as possible.
Therefore, it is reasonable to adopt the steering beam pattern that achieves the highest beam gain around the steering points.

Using the similar mathematical approach in \cite{wbgain}, we first investigate the beam gain approximation of the steering beam pattern.
The steering beam vector can be calculated as $\displaystyle \mathbf{w}_p = \mathbf{a} (\theta_p, r_p)$, which achieves highest beam gain at steering point $(\theta_p, r_p)$.
Then we have the following Lemma.
\begin{lemma} \label{lm:steergain}
  Let $\displaystyle \gamma_1=\frac{b}{\sqrt{2|a|}}$, $\displaystyle \gamma_2=\frac{\sqrt{2|a|}n_w}{2}$, where $b = \theta - \theta_p$, $\displaystyle a = \frac{d^2}{\lambda}\left\{\frac{1 - \theta_p^2}{r_p} - \frac{1 - \theta^2}{r}\right\}$.
  Then the beam gain of the steering beam pattern can be expressed as
  \begin{equation} \label{eq:steergain}
    \begin{split}
      &g(\mathbf{w}_p, \theta, r)\approx  \\ & \left|\frac{C(\gamma_1 + \gamma_2) - C(\gamma_1 - \gamma_2) + j(S(\gamma_1 + \gamma_2) - S(\gamma_1 - \gamma_2))}{2\gamma_2} \right|.
    \end{split}
  \end{equation}
\end{lemma}
The proof of \textbf{Lemma \ref{lm:steergain}} can be found in Appendix \ref{Ap:lemma1}.
Then we investigate the coverage region of the $l^rl^{\theta}$-th codeword $\mathbf{w}^p_{l^{\theta}l^r} = \mathbf{a} (\theta_{l^{\theta}l^r}, r_{l^{\theta}l^r})$ in the lower-layer codebook $\mathbf{W}_{n_{lv}}$.
The coverage region in $\mathbf{W}_{n_{lv}}$ is defined as $\displaystyle \mathcal{C}(\mathbf{w}^p_{l^rl^{\theta}}) = \left\{(\theta, r) \Big| g(\mathbf{w}_{l^{\theta}l^r}, \theta, r) \geq g(\mathbf{w}, \theta, r), \forall \mathbf{w} \in \mathbf{W}_{n_{lv}} \right\}.$
Based on (\ref{eq:steergain}), we can develop the boundary points set where target codeword $\mathbf{w}^p_{l^{\theta}l^r}$ achieves the same beam gain with its adjacent codewords. % $\mathbf{w}^p_{(l^{\theta}-1)l^r}$ and $\mathbf{w}^p_{l^{\theta}(l^r-1)}$.
%Specifically, when two steering points are located on the same distance ring \cite{channelest}, i.e., $\frac{1 - \theta_{l^{\theta}-1}^2}{r_{l^r}} = \frac{1 - \theta_{l^{\theta}}^2}{r_{l^r}}$,  we have $g(\mathbf{w}^p_{l^rl^{\theta}}, \theta, r) = g(\mathbf{w}^p_{l^r(l^{\theta}-1)}, \theta, r), \forall \left(\theta, r\right)$.
Specifically, when two steering points $(\theta_{p_1}, r_{p_1})$ and $(\theta_{p_2}, r_{p_2})$ are located on the same distance ring \cite{channelest}, i.e., $\frac{1 - \theta_{p_1}^2}{r_{p_1}} = \frac{1 - \theta_{p_2}^2}{r_{p_2}}$, the boundary points set can be expressed as $\mathcal{B}_{\theta} = \left\{(\theta, r) | \theta = \frac{\theta_{p_1} + \theta_{p_2}}{2}\right\}$.
Similarly, when two steering points $(\theta_{p_1}, r_{p_1})$ and $(\theta_{p_2}, r_{p_2})$ are located in the same direction, the boundary points set can be expressed as $\displaystyle \mathcal{B}_{r} = \left\{ (\theta, r) \Bigg| \frac{1-\theta^2}{r} = \frac{1-\theta_{p_1}^2}{2} \left(\frac{1}{r_{p_1}} + \frac{1}{r_{p_2}}\right) \right\}$.
Since the codeword $\mathbf{w}_{l^rl^{\theta}}$ achieves the highest beam gain within the boundary points set, the coverage region can be formulated in (\ref{eq:cv}) as shown on the top of this page.
\begin{figure*}[ht] %hb代表放在文章底部，%ht为放在文章顶部
  \centering
  \begin{equation} \label{eq:cv}
    \begin{split}
      \mathcal{C}(\mathbf{w}_{l^rl^{\theta}}) = \left\{ (\theta, r) \Bigg | \frac{\theta_{l^{\theta}} + \theta_{l^{\theta} - 1}}{2} \leq \theta < \frac{ \theta_{l^{\theta} + 1} + \theta_{l^{\theta}}}{2},
      \frac{1-\theta_{l^{\theta}}^2}{2} \left(\frac{1}{r_{l_{r}}} + \frac{1}{r_{l^r-1}}\right) \leq \frac{1-\theta^2}{r} < \frac{1-\theta_{l^{\theta}}^2}{2} \left(\frac{1}{r_{l^r+1}} + \frac{1}{r_{l^r}}\right) \right\}
    \end{split}
  \end{equation}
  \rule[0pt]{\textwidth}{0.02em}
\end{figure*}
Therefore, the steering points in $\mathbf{W}_{n_{lv}}$ can be sampled as $\displaystyle r_{l^r-1} = \frac{1}{l^r} \delta (1-\theta_{l^{\theta}}), l^r=1, 2, ..., n^l$ and $\displaystyle \theta_{l^{\theta}} =-1 + \frac{2l^{\theta}-1}{n^{\theta}}, l^{\theta}=1, 2, ..., n^{\theta}$, where $\delta$ is the scalar that controls the distance of different rings.

\begin{figure}
  \centering
  \includegraphics[width=0.45\textwidth]{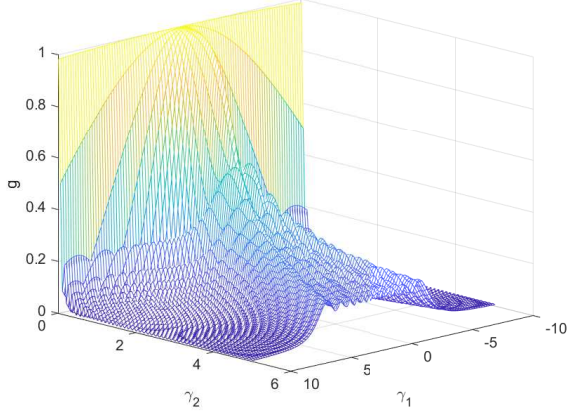}
  \caption{The approximated beam gain of steering vector in (\ref{eq:steergain}).}
\label{fig:steergain}
\end{figure}
As shown in Fig. \ref{fig:steergain}, the beam gain decreases monotonically with $|b|$ when $|a|$ is fixed.
Therefore, the minimum beam gain $g_{min}$ in the region $\mathcal{C}(\mathbf{w}_{l^rl^{\theta}})$ can be achieved at the corner of the coverage region.
%i.e. $\forall (\theta, r) \in \mathcal{C}(\mathbf{w}_{l^rl^{\theta}})$,
%\begin{equation}
%  \begin{split}
%    g\left(\mathbf{w}_{l^rl^{\theta}}, \frac{\theta_{l^{\theta}} + \theta_{l^{\theta} - 1}}{2}, \frac{2r_{l^r-1}r_{l^r}}{r_{l^r-1} + r_{l^r}}\right) \leq g(\mathbf{w}_{l^rl^{\theta}},\theta,r). \\
%  \end{split}
%\end{equation}
The lower-layer codebook design is to cover the whole Fresnel region by all codewords, i.e.,
\begin{equation} \label{eq:cover}
  \bigcup_{l^r=1}^{n^{l}} \bigcup_{l^{\theta}=1}^{n^{\theta}} \mathcal{C}(\mathbf{w}_{l^rl^{\theta}}) = [-1, 1] \times (r_{min}, + \infty).
\end{equation}
This coverage could be accomplished by adjusting $n_{\theta}$ and the scalar $\delta$ until the beam gain satisfies $g_{min} >= \rho$.

\subsection{Upper-layer Codebook Design}
The upper-layer codebook $\mathcal{W}^{u}$ is designed to reduce the beam searching steps.
In the upper-layer codebook, the $i$-th codeword of the $l$-th layer codebook $\mathbf{W}_{l}[i]$ is assigned to a subset of codeword $\mathcal{W}^i_{l+1} = \{\mathbf{W}_{l+1}[i_1], \mathbf{W}_{l+1}[i_2], ..., \mathbf{W}_{l+1}[i_k]\}$ in the next layer.
When the codeword $\mathbf{W}_{l}[i]$ is selected, the beam searching in the next level only needs to be conducted within the subset $\mathcal{W}^i_{l+1}$.
In this way, the beam searching steps can be greatly reduced compared with the exhaustive beam searching, especially when the number of codewords in the lower-layer codebook is large.
Inspired by the beam rotation method in \cite{cbbwmss}, we propose beam relocation and beam rotation methods to help the upper-layer codebook design in the near-field situation.

\emph{1) Beam Rotation}:
The beam gain of arbitrary beam vector $\mathbf{w}$ can be rotated through the following Lemma
\begin{lemma} \label{lm:rotation}
  For arbitrary beam $\mathbf{w}$, let $\tilde{\theta} = (\theta - \Delta \theta)$, $\displaystyle \tilde{r} = \frac{1-(\theta-\Delta \theta)^2}{1-\theta^2} r$. Then we have
  \begin{equation}\label{eq:ro}
    \begin{split}
    g(\mathbf{w} \circ \sqrt{n_w} \mathbf{a}(\Delta \theta, \infty), \theta, r) &= g(\mathbf{w}, \tilde{\theta} , \tilde{r}).
    \end{split}
  \end{equation}
\end{lemma}
The proof of \textbf{Lemma \ref{lm:rotation}} can be found in Appendix \ref{Ap:lemma2}.
It is noted that the target location $(\tilde{\theta} , \tilde{r})$ and the original location $(\theta, r)$ with the same beam gain satisfy $\displaystyle \frac{1-(\theta-\Delta \theta)^2}{\tilde{r}} = \frac{{1-\theta^2}}{r} $.
Equation (\ref{eq:ro}) provides the insights that by making Hadamard product with a far-field steering vector, the beam pattern is rotated along an elliptical trajectory.
It is also noted that the coverage of the beam pattern also rotates along the same trajectory.
The coverage rotation can be formulated as:
\begin{equation}
  \begin{split}
    \mathcal{C}(\tilde{\bm{w}}) &= \left\{(\theta, r)  \bigg| g(\mathbf{w} \circ \sqrt{n_w} \mathbf{a}(\Delta \theta, \infty), \theta, r) > \rho \right\} \\
    &= \left\{(\theta, r)  \bigg| g(\mathbf{w}, \tilde{\theta}, \tilde{r}) > \rho \right\}.
  \end{split}
\end{equation}

\begin{figure}
  \centering
  \includegraphics[width=0.45\textwidth]{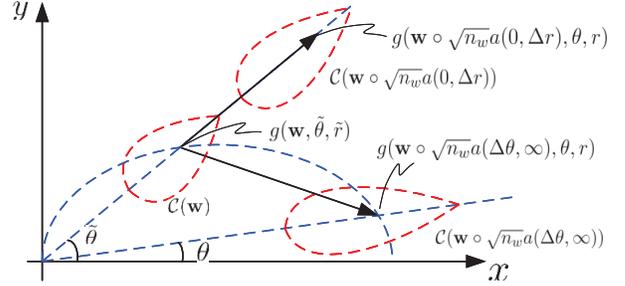}
  \caption{The Illustration of Beam Rotation and Beam Relocation.}
\label{fig:beamro}
\end{figure}

\emph{2) Beam Relocation}:
The beam gain of arbitrary beam vector $\mathbf{w}$ can be relocated through the following Lemma
\begin{lemma} \label{lm:relocation}
  For arbitrary beam $\mathbf{w}$, let $\tilde{\theta} = \theta$, $\displaystyle \frac{1}{\tilde{r}} = \frac{1}{r} - \frac{1}{\Delta r(1-\theta^2)}$.
  Then we have
  \begin{equation}
    \begin{split}
    g(\mathbf{w} \circ \sqrt{n_w} \mathbf{a}(0, \Delta r), \theta, r) &= g(\mathbf{w}, \tilde{\theta} , \tilde{r}).
    \end{split}
  \end{equation}
\end{lemma}
The proof of \textbf{Lemma \ref{lm:relocation}} is similar to the proof in Appendix \ref{Ap:lemma1} and is omitted due to space limit.
The beam gain relocation is done along the same direction, while the target relocation position $\tilde{r}$ is determined by the direction $\theta$ as well as $\Delta r$.

For the $l$-th layer hierarchical codebook, we can first generate one suitable beam pattern $\mathbf{w}^{ori}_l$ that covers areas around $(0, \infty)$.
Then we can use the beam relocation method to generate $n^{r}_l$ beams that cover $r$-dimension with direction $\theta = 0$.
Then the codewords can be rotated towards directions $\mathcal{D} = \{-1 + \frac{2n-1}{n^{\theta}_l}, n=1, 2, ..., n^{\theta}_l\}$.
In this way, an arbitrary initial beam pattern can be allocated to provide coverage for the whole Fresnel region.
The detailed codebook construction method is formulated in Algorithm 1.

\begin{breakablealgorithm}
  \renewcommand{\algorithmicrequire}{\textbf{Input:}}
  \renewcommand{\algorithmicensure}{\textbf{Output:}}
  \caption{Construction of Hierarchical Codebook in Near-field Region} \label{algo:reconstruct}
  \begin{algorithmic}[1]
  \REQUIRE Number of levels $n_{lv}$, initial beam pattern sets $\{\mathbf{w}^{ori}_1, \mathbf{w}^{ori}_2, ..., \mathbf{w}^{ori}_{n_{lv}} \}$
  \ENSURE Hierarchical codebook $\mathcal{W}$
  \FOR{$l = 1, 2, ..., n_{lv}$}
  \STATE Let $r_1 = \infty, i_r = 1$
  \WHILE{$r_{i_r} > r_{min}$}
    \STATE Select $r_{i_r}$ that satisfies coverage condition
    \STATE Beam relocation $\mathbf{\tilde{W}}_l[i_{r}] = \mathbf{w}^{ori}_l \circ \mathbf{a} (0, r_{i_r})$
    \FOR{$i_{\theta} = 1, 2, ..., n^{\theta}_l$}
        \STATE $\mathbf{W}_l[i_{r}, i_{\theta}] = \mathbf{\tilde{W}}_l[i_{r}] \circ \mathbf{a} \left(-1 + \frac{2i_{\theta}-1}{n^{\theta}_l}, \infty\right)$
    \ENDFOR
    \STATE $i_r = i_r + 1$.
  \ENDWHILE
  \ENDFOR
  \end{algorithmic}
\end{breakablealgorithm}

\section{Simulation Result}
In simulation setups, we consider a single BS equipped with $n_w$ antennas to provide beamforming for $n_u$ users randomly located in the Fresnel region.
The channels in (\ref{eq:channel}) between users and the BS consist of LOS paths.
The detailed parameters are summarized in Table. \ref{tb:set}.
\begin{table}
  \centering
  \caption{Communications Settings}\label{tb:set}
  \begin{tabular}{cc}
      \toprule
      \textbf{Parameters} & \textbf{Value}\\
      \hline
      Carrier Frequency $f$ & $40$ GHz\\
      Bandwidth $B$ & $1$ GHz\\
      Number of Elements $n_w$ & $256$ \\
      Number of Users $n_u$ & $100000$ \\
      Fresnel distance $r_{min}$ & $5.43m$ \\
      \bottomrule
  \end{tabular}
\end{table}

\begin{figure}
  \centering
  \includegraphics[width=0.45\textwidth]{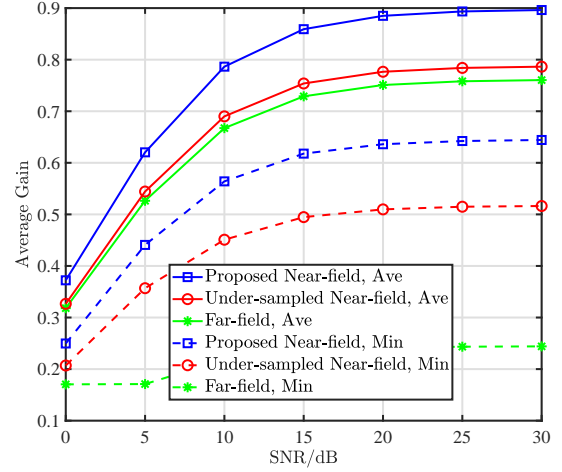}
  \caption{The simulation results of average and minimum beamforming gain}
\label{fig:gain}
\end{figure}

The minimum gain in the lower-layer codebook is set as $\rho = 0.64$ \cite{cbbwmss}.
The numbers of codewords in lower-layer codebook along $r$-dimension and $\theta$-dimension are $n^{\theta} = 512$ and $n^{r} = 5$, respectively.
The under-sampled codebook in \cite{channelest} is represented in the same way as the proposed lower-layer codebook, with $n^{\theta} = 256$ and $n^{r} = 4$.
As shown in Fig. \ref{fig:gain}, the proposed lower-layer codebook achieves $11.07$\% and $30.65$\% higher average and minimum beamforming gain than the under-sampled codebook when SNR is $20$dB, respectively.
This is because the lower-layer codebook is designed to satisfy the beam coverage condition in (\ref{eq:cover}).
Therefore, more codewords are assigned to fill the service hole in the under-sampled codebook.
It can also be observed that the far-field codebook has the worst performance for the average and minimum beamforming gain.
The average and minimum beamforming gain of the lower-layer codebook is $21.76$\% and $218.36$\% higher than the far-field codebook, respectively.
This is because the spherical wavefront characteristic of the near-field situation is not negligible.

Then we conduct the simulation of the upper-layer codebook.
The number of levels is set as $n_{lv} = 9$.
The number of codeword in the $l$-th layer codebook along $\theta$ dimension is set as $n^{\theta}_l = 2^{l}$, while $r_{i_r}$ is selected as $g(\mathbf{\tilde{W}}_l[i_{r}], r_{i_r}, 0) = 0.5 g(\mathbf{w}^{ori}_l, \infty, 0)$.
We test 3 different initial beam patterns:\\
$\bullet$ \emph{Deact} \cite{cbdeact}: The wider codewords are generated by turning off antennas in ULA.\\
%The Deact codeword can be expressed as
%\begin{equation}\label{eq:deact}
%  \begin{split}
%     \mathbf{w}^{de}_{l} &=  \left[\mathbf{0}^T_{\frac{n_w-n_a}{2}}, \mathbf{a}^T_{n_a}(0, \infty), \mathbf{0}^T_{\frac{n_w-n_a}{2}} \right]^T,
%  \end{split}
%\end{equation}
%where $n_a = n_w / 2^{(n_{lv}-l)}$ is the number of active elements in the ULA.\\
$\bullet$ \emph{BMW-SS} \cite{cbbwmss}: The beam pattern is suited for far-field hierarchical codebook using joint deact and sub-array approaches. \\
$\bullet$ \emph{Quadric} \cite{quadbeam}: A wide beam pattern that is generated using a quadratic phase.

\begin{figure}
  \centering
  \includegraphics[width=0.45\textwidth]{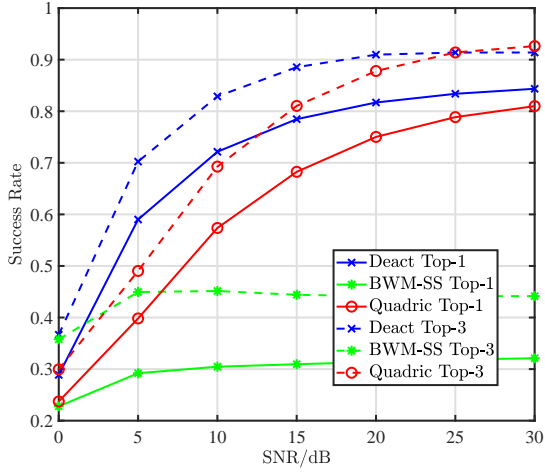}
  \caption{The success rate of different codebooks}
\label{fig:success_rate}
\end{figure}

It can be seen in Table. \ref{tb:steps} that the proposed hierarchical codebook design greatly reduces the total number of searching steps compared with exhaustive search in the lower-layer codebook.
We also conduct simulations to investigate the Top-$k$ success rate of different upper-layer beam patterns, as shown in Fig. \ref{fig:success_rate}.
It can be seen that the \emph{Deact} codebook achieves the highest success rate in both Top-1 and Top-3 situations.
This phenomenon can be explained as that the \emph{Deact} beam pattern in the upper-layer codebook achieves the highest gain at the target location $(r_{i_{r}}, \theta_{i_{\theta}})$, while the beams in other two upper-layer codebooks do not achieve the highest beam gain at the target locations.
Therefore, there would be a wrong selection of the beam index in \emph{BMW-SS} and \emph{Quadric} upper-layer codebook.
Since \emph{Deact} beam pattern has the lowest average searching steps and achieves the best success rate, it is the optimal initial beam pattern.

\begin{table}
  \centering
  \caption{Average Searching Steps}\label{tb:steps}
  \begin{tabular}{|c|c|c|c|c|}
      \hline
      Codebook & Deact & BMW-SS & Quadric & Exhaustive\\
      \hline
      Steps & 18.60 & 20.43 & 22.08 & 2560\\
      \hline
  \end{tabular}
\end{table}

\section{Conclusion}
In this paper, we proposed a hierarchical codebook design method that is suitable for the near-field situation.
We first formulated the approximation beam gain of the steering beam pattern and designed the lower-layer beam codebook that satisfies the beam coverage condition of the Fresnel region.
Then, we developed the beam rotation and beam relocation methods to place the beam coverage in the upper-layer codebook at proper locations.
The simulation results show that the proposed lower-layer codebook can achieve higher average and minimum beamforming gain compared with traditional far-field codebooks.
Also, the hierarchical codebook can greatly reduce the beam training steps compared with the exhaustive searching.
\appendices
\section{Proof of Lemma \ref{lm:steergain}} \label{Ap:lemma1}
%^ Let $b = \theta - \theta_p$, $\displaystyle a = \frac{d^2}{\lambda}\left\{\frac{1 - \theta_p^2}{r_p} - \frac{1 - \theta^2}{r}\right\}$.
The beam gain of the steering beam pattern at location $(\theta, r)$ can be calculated as
  \begin{equation}
    \begin{split}
      g(\mathbf{w}, \theta, r) &\approx \left|\frac{1}{n_w} \sum_{-(n_w-1)/2}^{(n_w-1)/2} e^{j\pi (an^2+bn)}\right| \\
      &\approx \left|\frac{1}{n_w} \int_{-n_w/2}^{n_w/2}   e^{j\pi (an^2+bn) } dn\right| \\
      &= \left|\frac{1}{n_w} \int_{-n_w/2}^{n_w/2}   e^{j\pi a[(n+\frac{b}{2a})^2 - (\frac{b}{2a})^2] } dn\right|. \\
    \end{split}
  \end{equation}
  When $a \geq 0$, let $\displaystyle t=\sqrt{2a}\left(n+\frac{b}{2a}\right)$, and then we have
  \begin{equation}
    \begin{split}
      g(\mathbf{w}, \theta, r) &\approx \left|\frac{1}{n_w e^{j\pi \frac{b^2}{2a}}} \int_{\sqrt{2a}(-\frac{n_w}{2} + \frac{b}{2a})}^{\sqrt{2a}(\frac{n_w}{2} + \frac{b}{2a})}   e^{j\frac{\pi}{2} t^2 } d\left(\frac{t}{\sqrt{2a}}\right)\right|. \\
    \end{split}
  \end{equation}
  Let $\displaystyle \gamma_1=\frac{b}{\sqrt{2a}}$, $\displaystyle \gamma_2=\frac{\sqrt{2a}n_w}{2}$. We have
  \begin{equation}
    \begin{split}
      &g(\mathbf{w}, \theta, r)\approx F(\gamma_1, \gamma_2) \\ & = \left|\frac{C(\gamma_1 + \gamma_2) - C(\gamma_1 - \gamma_2) + j(S(\gamma_1 + \gamma_2) - S(\gamma_1 - \gamma_2))}{2\gamma_2} \right|
    \end{split}
  \end{equation}
  where $C(\gamma) = \int_{0}^{\gamma} \cos(\frac{\pi}{2}t^2)dt$ and $S(\gamma) = \int_{0}^{\gamma} \sin(\frac{\pi}{2}t^2)dt$.
  Similarly, when $a < 0$, $\displaystyle \gamma_1=\frac{b}{\sqrt{-2a}}$, $\displaystyle \gamma_2=\frac{\sqrt{-2a}n_w}{2}$, we have $g(\mathbf{w}, \theta, r)\approx |F^*(\gamma_1, \gamma_2)| = |F(\gamma_1, \gamma_2)|$.
  In this way, \textbf{Lemma \ref{lm:steergain}} is proved.

  \section{Proof of Lemma \ref{lm:rotation}} \label{Ap:lemma2}
  The beam gain of vector $\mathbf{w}$ can be calculated as
    \begin{equation}
      \begin{split}
      g(\mathbf{w}, \tilde{\theta} , \tilde{r}) &\approx \left|\frac{1}{n_w} \sum_{-(n_w-1)/2}^{(n_w-1)/2} [\mathbf{w}]_n e^{jn\pi \tilde{\theta} - j\frac{2\pi}{\lambda}
        n^2 d^2\left(\frac{1 - \tilde{\theta}^2}{2 \tilde{r}}\right)}\right| \\
      &= \left|\frac{1}{n_w} \sum_{-(n_w-1)/2}^{(n_w-1)/2} [\mathbf{w}]_n e^{jn\pi (\theta - \Delta \theta) - j\frac{2\pi}{\lambda}
      n^2 d^2\left(\frac{1 - \theta^2}{2 r}\right)}\right| \\
      &= |\mathbf{a}(\theta, r)^H \left[\mathbf{w} \circ \sqrt{n_w} \mathbf{a}(\Delta \theta, \infty)\right]| \\
      &= g(\mathbf{w} \circ \sqrt{n_w} \mathbf{a}(\Delta \theta, \infty), \theta, r)
      \end{split}
    \end{equation}
    Then \textbf{Lemma \ref{lm:rotation}} is proved.
\bibliographystyle{IEEEtran}
\bibliography{ref}

\end{document}